\documentclass{article}
\usepackage{spconf,amsmath,graphicx,multirow}
\title{Convolutional neural networks for passive monitoring of a shallow water environment using a single sensor}
\twoauthors
  {\begin{tabular}{c}Eric L. Ferguson\sthanks{Work supported by Defence Science and Technology Group Australia and IEEE Oceanic Engineering Society Scholarships.}, Rishi Ramakrishnan, \\ Stefan B. Williams\end{tabular}}
	{Australian Centre for Field Robotics\\
		 The University of Sydney, Australia}
  {Craig T. Jin}
	{Computing and Audio Research Laboratory\\
		 The University of Sydney, Australia}

\begin{document}
\ninept
\maketitle
\begin{abstract}
A cost effective approach to remote monitoring of protected areas such as marine reserves and restricted naval waters is to use passive sonar to detect, classify, localize, and track marine vessel activity (including small boats and autonomous underwater vehicles). Cepstral analysis of underwater acoustic data enables the time delay between the direct path arrival and the first multipath arrival to be measured, which in turn enables estimation of the instantaneous range of the source (a small boat). However, this conventional method is limited to ranges where the Lloyd's mirror effect (interference pattern formed between the direct and first multipath arrivals) is discernible. This paper proposes the use of convolutional neural networks (CNNs) for the joint detection and ranging of broadband acoustic noise sources such as marine vessels in conjunction with a data augmentation approach for improving network performance in varied signal-to-noise ratio (SNR) situations. Performance is compared with a conventional passive sonar ranging method for monitoring marine vessel activity using real data from a single hydrophone mounted above the sea floor. It is shown that CNNs operating on cepstrum data are able to detect the presence and estimate the range of transiting vessels at greater distances than the conventional method.
\end{abstract}
\section{Introduction}
\label{sec:intro}
Despite the long-term usage of traditional passive acoustics for sound-source localization, poor performance persists in some scenarios. Current conventional, single-sensor source localization methods are limited in their effective range, which is further degraded in low SNR situations.
Time delay estimation aims to measure the time difference of arrival (TDOA) between propagation paths of an acoustic signal and is a fundamental approach for classifying, localizing and tracking sources of radiated acoustic noise. A common approach to the passive ranging of a sound source is to measure the TDOA of a signal at multiple, spatially distributed receivers~\cite{Carter1981, Carter1993, Chan1994, Benesty2004}.
The TDOA measured between two coherent signal arrivals at a single receiver is geometrically equivalent to the TDOA measured by a single arrival propagating to two vertically-spaced receivers~\cite{Hamilton1992}.
Passive acoustic ranging using a single sensor is achieved by measuring the TDOA of an acoustic signal as it arrives via direct and indirect underwater sound propagation paths. For example, the TDOA between the direct path signal and the multipath signal can be used to yield the instanenous range of the acoustic source~\cite{Ferguson2005}.
Passive acoustic ranging using a single sensor facilitates deployment, lowers hardware costs, and minimizes the equipment footprint when compared with multi-sensor arrays.

The acoustic characteristics of a shallow water environment such as a harbour or port are variable in both space and time with high levels of clutter, background noise, and multipath reflection. 
Time delay estimation by cepstral analysis is able to outperform other methods (such as autocorrelation analysis) in these scenarios~\cite{Gao2008}, however this method is limited to ranges where the Lloyd's mirror effect is discernible, i.e. only at short ranges and when the SNR of the recorded source is sufficiently high.

A CNN is proposed that operates on cepstral inputs to detect and range an acoustic source passively in a shallow water environment.
The CNN based implementation has an important advantage over other methods in that the TDOA information for more complex multipaths can be exploited, rather than peak quefrency values used in conventional methods.
This increases the range at which source tracking is possible. By considering additional propagation paths such as paths with two or more boundary reflections, it is hypothesized that the source range can be estimated at greater distances, even when the Lloyd's mirror interference pattern is not discernible by a human observer. 
The CNNs are trained using real, single channel acoustic recordings of a surface vessel under way in a shallow water environment.
CNNs operating on both cepstrum and cepstrogram inputs are considered and their performances compared.
The proposed models are shown to detect and range sources successfully at greater distances and in varied SNR situations and are compared with a conventional single-sensor passive sonar localization method.
Generalization performance of the network is tested by ranging another, previously unseen vessel with different radiated noise characteristics.
To the best of our knowledge, this is the first acoustic localization network to utilize the TDOA information in a reverberant environment to range and detect a source passively with just one sensor.

The contributions of this work are:
\begin{itemize}
\item Development of a CNN for the passive ranging of acoustic broadband noise sources in shallow water environment at greater distances than conventional methods allow;
\item Cepstral liftering of network inputs to improve ranging of other radiated noise sources;
\item Data augmentation technique where colored noise is added to training data to improve robustness in varied SNR scenarios; and
\item A unified, end-to-end network for the joint detection and ranging of acoustic sources.
\end{itemize}

\section{Detection and Ranging CNN}
\label{sec:Contributions}
A neural network is a machine learning technique that maps the input data to a label or continuous value, through a multi-layer non-linear archictecture and has been successfully applied to applications such as image/object classification \cite{Krizhevsky2012,Girshick2014} and terrain classification using acoustic sensors\cite{Valada2015}. CNNs learn sets of filters that span small regions of the input data, enabling them to learn local correlations.

\subsection{Architecture}

\begin{figure}[t]
  \centering
\includegraphics[width=0.48\textwidth, clip=true, trim=65 0 55 0 ]{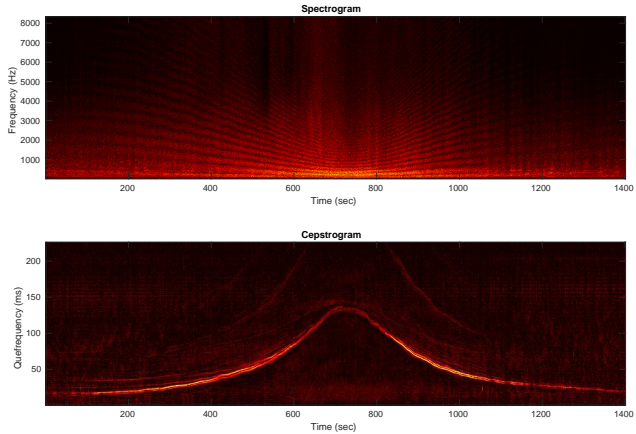}
 \caption{\textbf{a)} Spectrogram showing the Lloyd's mirror for a surface vessel as it transits over a hydrophone at close range, and 
 \textbf{b)} the corresponding cepstrogram }
\label{fig:specandcepst}
\end{figure}

Since an acoustic source has an effect on the cepstrum, it is possible to create a unified network for classifying the presence/absence of a vessel, and determining the range of the detected vessel.
The network structure is as follows: The first layer consists of $48$ convolutional filters of size $10 \times n$, where $n$ refers to the input width, as is discussed further in Section~\ref{sec:expResinput}.
Both the second and third layers consist of $48$ convolutional filters of size $10 \times 1$.
The third layer is then an input layer to a fully connected hidden layer of 200 neurons with a single regression output and a binary softmax classification output.
All layers (excluding output layers) use rectified linear units as activation functions.
Since resolution is important for the accurate ranging of an acoustic source, max pooling is not used in the network's architecture.

\subsubsection{Input}
~\label{sec:cepts}
A cepstrum can be derived from various spectra such as the complex or differential spectrum. 
For the current approach, the power cepstrum (referred to in this paper as the cepstrum) is used and is derived from the power spectrum of a recorded signal. 
Cepstral analysis is based on the principle that the logarithm of the power spectrum for a signal containing echoes has an additive periodic component due to the echoes from multipath reflections~\cite{Lo2003}. 
This additive periodic component is evident when examining the Lloyd's mirror effect in the spectrogram when an acoustic source travels past the hydrophone at close range as seen in  Fig.~\ref{fig:specandcepst}a).
The cepstral representation of the signal is neither in the time, nor frequency domain but rather it is in the quefrency domain~\cite{Bogert1963}.
Where the original time waveform contained an echo the cepstrum will contain a peak and thus the TDOA between propagation paths of an acoustic signal can be measured by examining peaks in the cepstrum~\cite{Oppenheim2004}. The cepstrogram (an ensemble of cepstrum as they vary in time) is shown in Fig.~\ref{fig:specandcepst}b). 

The cepstrum $\hat{x}(n)$ is obtained by the inverse Fourier transform:
\begin{equation}
 \label{eq:cepst1}
  	\hat{x}(n) = F^{-1}\big(log|S(f)|^2\big),
\end{equation}
where $S(f)$ is the Fourier transform of a discrete time signal $x(n)$.

In order to detect and range a source using a single sensor, information about the time delay between signal propagation paths is required. Although such information is contained in the raw signals, it is beneficial to represent it in a way that can be learnt by the network easily.
There are several ways to represent time delay information. Motivated by work in \cite{Gao2008}, the cepstrum is chosen as network input, since it provides TDOA information between signal propagation paths that can be used to passively range the vessel.
The capability of cepstrum analysis in extracting TDOA information is superior to other methods (such as autocorrelation) in the presence of multipath reflections and strong transients found in a shallow water environment~\cite{Gao2008}.

The first layer's convolutional filter spans the entire input width in order to average neighbouring cepstral values and reduce the impact of shot noise and other short-duration clutter.
By using filters that span the entire width of the input, networks can be robust to short-duration changes in the cepstrogram.
The temporal difference of cepstra in the cepstrogram is not important for the task at hand since for the present experiments only the instantaneous range and detection is of interest.

%

\subsubsection{Output}
For each input into the network, the network classifies the presence or absence of a vessel using binary softmax classification. If the vessel is present, the range of the acoustic source is predicted with a regression output. 

\subsubsection{Cesptral Liftering}
\label{sec:liftering}
For a given source-sensor geometry, there is a finite bounded range of possible TDOA values. 
Distant acoustic sources will have TDOA values that tend to zero and as the source-sensor separation distance decreases the TDOA values will tend to a maximum value. 
TDOA values greater than this geometry dependant maximum are not useful for the passive sonar ranging problem, hence upper bounds of the cepstrum can be discarded.

Cepstrum values near zero mostly contain pitch information for the broadband noise source, and not TDOA information for different signal propagation paths. Acoustic sources of interest are varied in their radiated noise characteristics; for example, the inception of propeller cavitation leads to a significant increase in the intensity and bandwidth of the radiated noise. For this reason, lower quefrency values are likely to be highly source dependant and are thus not useful for the passive sonar ranging problem. Hence lower bounds of the cepstrum can be discarded.

Similar to filtering in the frequency domain by windowing a spectral represenation of a signal, liftering involves linear filtering of the log spectrum (in the quefrency domain) by windowing~\cite{Bogert1963}. Only quefrencies between some range contain useful TDOA information for passive acoustic ranging, as described above. The cepstrum can be liftered (filtered in quefrency) to remove information not useful for passive ranging of the source.
This has the added benefit of reducing computational complexity for forward and backward propagation through a network, since input dimensions are smaller and fewer convolutional filters are required.

\subsection{Data Augmentation}
\label{sec:dataAug}
The acoustic noise characteristics of a shallow water environment is variable in both space and time with high levels of clutter, background noise and multipath reflection. For example, different times of day have varying levels of biological noise.
Further, acoustic sources vary in the level of sound power they emit. For robust ranging and detection of other sources it is important for the network to be invariant to changes in radiated or background noise levels.
By performing transformations to recorded signals the number of training examples is increased and network develops invariance to particular signal variations.

Since acoustic classification can be strongly affected by environmental noise, Valada~\cite{Valada2015} et. al shows that by augmenting raw acoustic data with additive white Gaussian noise, classification performance can increase in degraded SNR situations. 
This paper proposes augmenting raw acoustic data by adding colored noise with the same power spectral density (PSD) as background noise recordings during network training.
The PSD is taken from background noise recorded by the same hydrophone when no surface vessel is present.
Adding colored noise with the same PSD as background noise recordings simulates situations with either a quiet source or high levels of background noise.
By injecting colored noise to training examples the CNN performance can be improved by increasing robustness to SNR variations.
Furthermore, when $n>1$ training examples can be flipped along the quefrency axis, providing additional training examples.

\subsection{Joint Training}
The objective of the network is to predict the presence or absence of an acoustic source from reverberant and noisy single-channel input signals. If the source is present, then the range relative to the hydrophone is predicted.
Previously, it was found that ranging the vessel was a more difficult problem for the CNN and required more hidden units than vessel detection~\cite{Ferguson2016}.
This is to be expected since ranging is dependent on the location of cepstral features, whereas detection is only dependent on the presence of them.
The total objective function  $E$ minimized during network training is given by the weighted sum of the ranging regression loss $E_r$ and the detection loss $E_d$, such that:
\begin{equation}
E = \alpha E_{d} + (1-\alpha) E_{r} ,
\end{equation}
where $E_{r}$ is the L$_1$ norm and $E_{d}$ is the log loss over two classes. The two terms are weighted by parameter $\alpha$.
Training is performed by initially setting $\alpha = 0$, such that only the regression term is significant. Training is stopped when validation error does not decrease appreciably per epoch. 
Subsequently, due to the magnitude difference between $E_r$ and $E_d$, $\alpha $ is set to $0.99$ during joint training. Training is stopped when the validation error did not decrease appreciably per epoch.
For training data with no vessel present, there was no range label and $E_r$ was ignored, i.e. gradients obtained from the regression output for training samples with no boat were masked out.
In order to further prevent overfitting, regularization through dropout \cite{Srivastava2014} is used at the final, fully connected layer when training. A dropout rate of 50\% is used.

\section{Experimental Results}
\label{sec:experiment}

Passive ranging on a transiting vessel was conducted using a single-sensor algorithmic method described in \cite{Ferguson2005}, and CNNs with both cepstrum ($n=1$) and cepstrogram ($n=8$) inputs.
Their effectiveness is compared.
Generalization of the CNNs is also demonstrated by detecting and ranging an additional, unseen vessel with different radiated noise and SNR characteristics.

\subsection{Dataset}
Acoustic data of a motorised boat transiting in a shallow water environment over a hydrophone were recorded at a sampling rate of $250$~kHz.
Recordings start when the vessel is up to $500$~m away from the sensor. The vessel then transits over the hydrophone and recording is terminated when the vessel is $500$~m away.
The boat was equipped with a DGPS tracker, which logged its position relative to the recording hydrophone at $0.1$~s intervals.
28 transits were recorded over a two day period. Background noise was also recorded when there was no vessel present, over the same period.
20,000 training examples were randomly chosen, with an equal number of vessel transit recordings and background noise recordings. 
A further 5,000 labelled examples were reserved for CNN training validation.
The recordings were preprocessed as outlined in Section~\ref{sec:cepts}, ~\ref{sec:liftering} and~\ref{sec:dataAug}.
The networks are implemented in MatConvNet and are trained with stochastic gradient descent using a NVIDIA GeForce GTX 770 GPU. Due to GPU memory limitations, the gradient descent was calculated in batches of 256 training examples. The networks were trained with a learning rate of $1\times 10^{-6}$, weight decay of $5\times 10^{-4}$ and momentum of $0.9$. 

Additional recordings of the vessel were used to measure the performance of the methods. These recordings are referred to as the test dataset and contain $4032$ labelled examples.

Additional acoustic data were recorded on a different date, using a different boat with different radiated noise characteristics. Acoustic recordings started when the transiting vessel was $300$~m away from the hydrophone, record the transit over the hydrophone, and end when the vessel is $300$~m away. This dataset is referred to as the generalization set and contains $7923$ labelled examples.

\subsection{Input of Network}
\label{sec:expResinput}
Cepstral features were used as input to the CNN. The cepstral features have a dimension of $m$~x~$n$, where $m$ is the number of quefrency bins in each cepstrum realization and $n$ is the input width of the cepstrogram, and is computed as follows. 
For every training example, the data was further subdivided into $n$ sections and the cepstrum values calculated for each section.
For each calculated cepstrum, only some range of quefrencies contain relevant TDOA information and are retained since the rest of the values are not useful for the task here - see Section \ref{sec:liftering}.
Cepstrum values more than $1.4$~ms are discarded since the shallow water environment geometry makes it unlikely that useful TDOA information is present. 
Cepstrum values less than $84$~$\mu$s are discarded, since they mostly contain source dependant pitch information. 
Thus, each cepstrogram input is liftered and samples $21$ through $350$ are used as input to the network only. This results in a $330$~x~$n$ input size, since $m=330$.
Colored noise was added to the recordings to change the SNR randomly between $-10$~dB to $50$~dB when training, as described in Section~\ref{sec:dataAug}.

Multiple CNNs with variable input widths were produced and their performances compared. 
The $n=1$ and $n=8$ CNNs are compared in the following section.
For $n=1$, a single realisation of the cepstrum is used. For $n=8$, an ensemble of cepstrum (or cepstrogram) is used.

\subsection{Comparison of Range and Detection Methods}
\begin{figure}[t]
\centering
\includegraphics[width=0.48\textwidth, clip=true, trim=20 0 45 0 ]{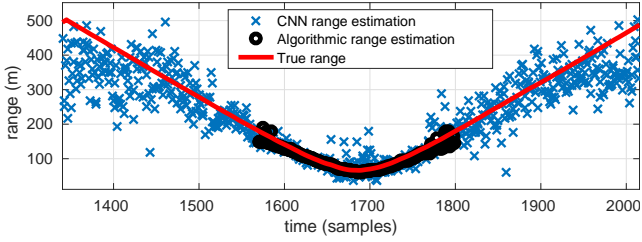}
\caption{A comparison of the two ranging methods, as they range a transiting vessel over time. The CNN range prediction refers to the estimated range given by the `$n=8$, with data aug' network. The true range shows the range of the vessel relative to the hydrophone, measured by the DGPS.}
\label{fig:results1}
\end{figure}

\begin{table}[]
\centering
\label{results:table1}
\begin{tabular}{|c|c|l|l|l}
\hline
\textbf{Network Input Width}                     & \multicolumn{2}{c|}{\textbf{n=1}}                         & \multicolumn{2}{c|}{\textbf{n=8}}               \\ \hline
\multicolumn{1}{|l|}{\textbf{Data Augmentation}} & \multicolumn{1}{l|}{no} & yes           & no & \multicolumn{1}{l|}{yes} \\ \hline
\textbf{Average Precision}                           & 0.9927                                & \multicolumn{1}{c|}{0.9942} & 0.9928           & \multicolumn{1}{l|}{\textbf{0.9978}}            \\ \hline
\end{tabular}
\caption{Comparison of detection performance for CNNs against the test dataset.}
\end{table}

\begin{figure}[t]
\centering
\includegraphics[width=0.48\textwidth, clip=true, trim=35 0 45 0 ]{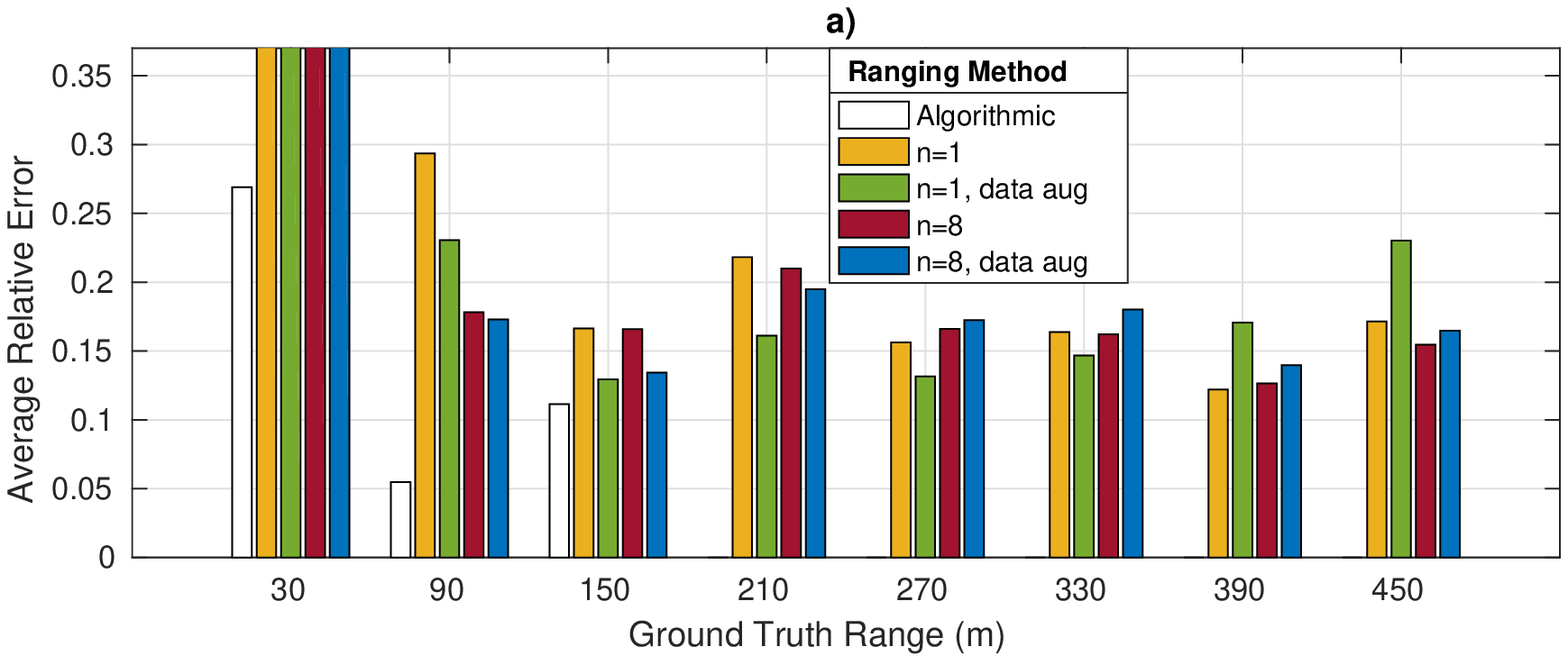}
\includegraphics[width=0.48\textwidth, clip=true, trim=35 0 45 0 ]{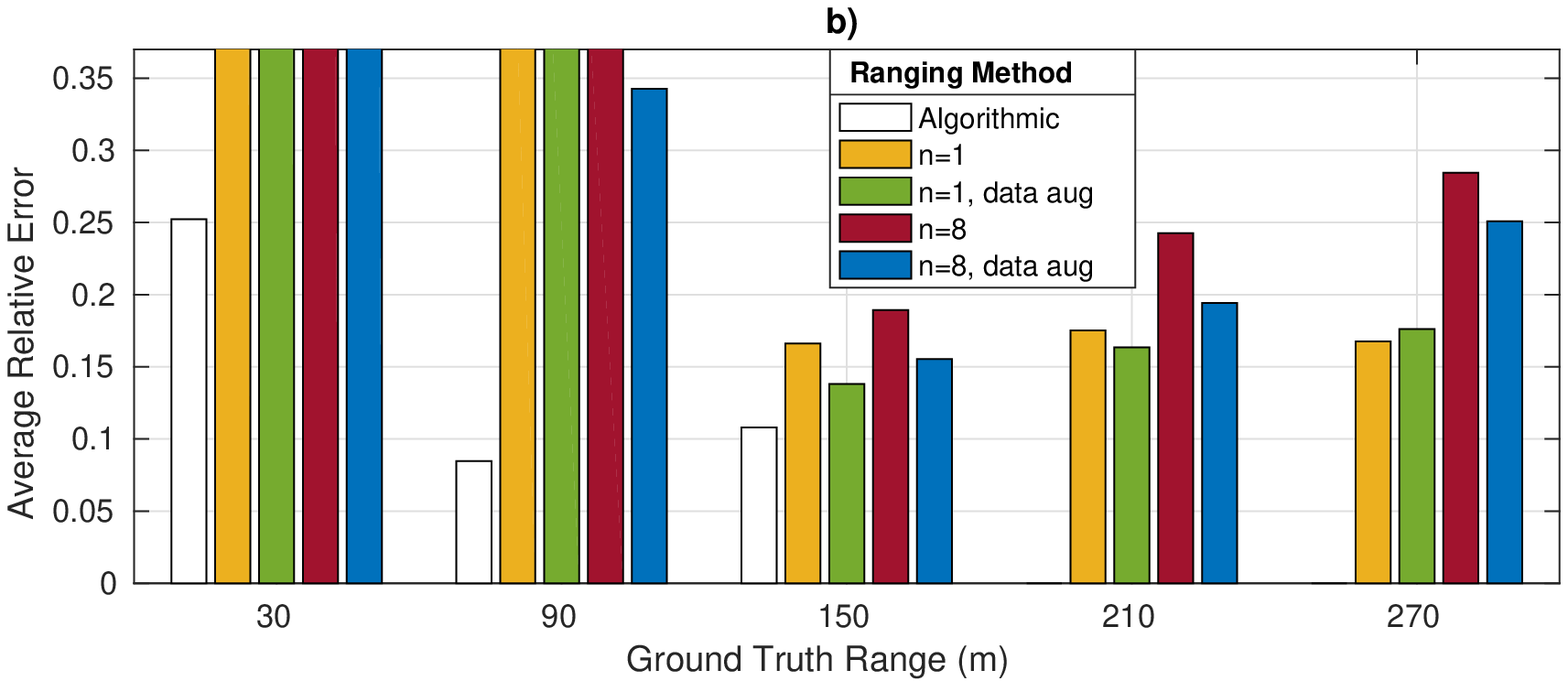}
\caption{Comparison of range estimation performance as a function of the vessels true range. It is not possible to determine the range of a vessel past $180$~m using conventional algorithmic methods, since the Lloyd's mirror interference pattern is not discernible. a) shows the performance when estimating the vessel's range in the test dataset. b) shows the performance when estimating the vessel's range in the generalization dataset.}
\label{fig:perfOverRange}
\end{figure}
\begin{figure}[t]
\centering
\includegraphics[width=0.48\textwidth, clip=true, trim=25 0 45 0 ]{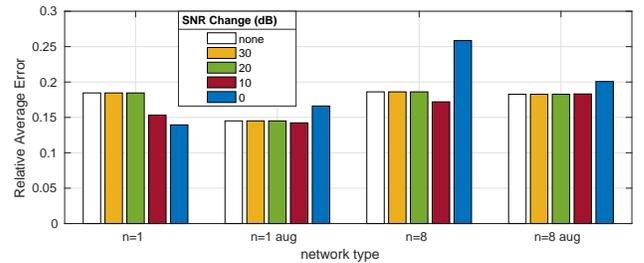}
\caption{Comparison of far field ($<180$~m) range estimation performance as a function of SNR.}
\label{fig:perfOverSNR}
\end{figure}

Algorithmic single sensor passive ranging was conducted, using the methods outlined in \cite{Ferguson2005}, where the TDOA values are measured by examining peaks in the cepstrum.
Fig.~\ref{fig:results1} compares algorithmic and CNN ranging over time for a vessel in transit. The algorithmic method is shown to successfully range a transiting vessel at ranges where the Lloyd's mirror interference pattern is present.
The CNN is shown to provide an estimate of the vessel range throughout the entire transit.

Table~\ref{results:table1} shows the average precision for each network when operating on the test dataset. Additive colored noise data augmentation improved CNN detection precision. Increasing network input width $n$ also improved the detection precision.

Fig.~\ref{fig:perfOverRange}~a) shows the performance of ranging methods as a function of the true range of the vessel for the test dataset. 
Fig.~\ref{fig:perfOverRange}~b) shows the performance of ranging methods as a function of the true range of the vessel for the generalization dataset. 
In the near field (ranges $<180$~m), the algorithmic ranging method out performs CNN ranging methods, achieving less average relative error.
CNN methods suffer from a significant bias in range estimates in the near field.
At source ranges further than $180$~m the algorithmic method fails completely and CNN methods are able to successfully estimate the range of the vessel.
The CNN is able to range the new vessel in the generalization set with a small impact to performance at these ranges.

Fig.~\ref{fig:perfOverSNR} shows the far field performance of the CNNs in estimating the vessels range under different SNR conditions. Test data was augmented with varying levels of colored noise, as described in Section~\ref{sec:dataAug}. For the $n=1$ case, data augmentation improved ranging performance in most cases. For the $n=8$ case, additive colored noise data augmentation improved ranging performance when the SNR was changed to $0$~dB only.

\section{Conclusions}
\label{sec:concl}
In this paper we introduce the use of a CNN for the detection and ranging of surface vessels in a shallow water environment.
Using liftered cepstra as input, the CNN detects the presence of a vessel and estimates its range relative to the recording hydrophone.
Several CNN architectures are evaluated.
A novel data augmentation technique is introduced, where colored noise of a similar PSD to recorded background noise is added to raw acoustic data when training.
This data augmentation improves performance in both vessel ranging and detection in some SNR scenarios.
Whilst the CNNs are outperformed by a conventional algorithmic method at short ranges ($<180$~m), the CNNs are able to estimate the vessel's range at further distances even when the Lloyd's mirror interference pattern is not easily identified.
The CNNs are robust to changes in the SNR and broadband spectral characteristics of marine vessels due to cepstral liftering of network inputs and novel data augmentation methods applied during network training.


\vfill\pagebreak
\bibliographystyle{IEEEbib}
\bibliography{strings,refs,library_RR}

\end{document}